\documentclass[prb,twocolumn,showpacs,preprintnumbers,amsmath,amssymb]{revtex4}

\usepackage{epsfig}
\usepackage{graphicx}
\usepackage{dcolumn}
\usepackage{bm}
\newcommand{\w}{\omega}
\newcommand{\tw}{\tilde \omega}

\renewcommand{\k}{{\bf k}}
\newcommand{\q}{{\bf q}}

\def\gl{\lower.35em\hbox{$\stackrel{\textstyle>}{\textstyle<}$}}
\begin{document}

\title{Dynamical polarizability of graphene beyond the Dirac cone approximation}
\author{T.~Stauber$^{1,2}$, J. Schliemann$^{3}$, and N.~M.~R.~Peres$^{1}$}

\affiliation{$^1$Centro de F\'{\i}sica  e  Departamento de
F\'{\i}sica, Universidade do Minho, P-4710-057, Braga, Portugal}
\affiliation{$^2$Dep. de F\'{\i}sica de la Materia Condensada, Universidad Aut\'{\o}noma de Madrid, E-28049 Madrid, Spain}
\affiliation{$^3$Institute for Theoretical Physics, University of Regensburg, D-93040 Regensburg, Germany}
\date{\today}

\begin{abstract}
We compute the dynamical polarizability of graphene beyond the usual Dirac
cone approximation, integrating over the full Brillouin zone. We find deviations at $\hbar\omega=2t$ ($t$ the hopping parameter) which amount to a logarithmic singularity due to the van Hove singularity and derive an approximate analytical expression. Also at low energies, we find deviations from the results obtained from the Dirac cone approximation which manifest themselves in a peak spitting at arbitrary direction of the incoming wave vector $\q$. Consequences for the plasmon spectrum are discussed.  
\end{abstract}

\maketitle
\section{Introduction}
Graphene is a novel two-dimensional system with many outstanding mechanical and electronic properties \cite{Geim09}. Especially the early observation of the ambipolar field effect\cite{Nov04} and of the odd integer quantum Hall effect \cite{Nov05Kim05}  have stimulated enormous research on the electronic structure of graphene. Only recently, the fractional quantum Hall effect was seen in suspended graphene\cite{Andrei09}. For a review of this newly emerging branch of condensed matter physics, see Ref. \cite{Neto09}.

To understand the unusual electronic properties of graphene, it often suffices to discuss the charge susceptibility. The static polarizability at $k_F$, e.g., gives the Thomas-Fermi screening length, important for transport properties \cite{Ando06,Adam07,StauberPeresGuinea07} whereas the dynamical polarizability at zero-wave number can explain the phonon softening \cite{Neto07} at the $\Gamma$-point. It is also used for the understanding of structural inhomogeneities in graphene, so-called ripples\cite{Gazit09} and the van der Waals interaction between graphene layers\cite{GomezSantos09}. 

For neutral graphene, the polarizability at zero temperature was first calculated by Gonzalez et al. \cite{Gon94}, the effect of temperature was discussed by Vafek \cite{Vafek06} and vertex corrections were considered in Ref. \cite{Mishchenko08}. For a gated system with finite chemical potential, the first expressions were given by Shung \cite{Shung86} in the context of graphite
and later by Wunsch et al. \cite{Wunsch06} and Hwang and Das Sarma \cite{Hwang07}. Also the extension to finite temperature has been performed, even though a closed analytical expression is then - as in the neutral case - not possible, anymore \cite{Ramezanali09}. Recently, the polarizability was discuss in the presence of a magnetic field \cite{Roldan09} and gapped graphene \cite{Pyatkovskiy09,Pedersen09}.

All these results originate from the Dirac cone approximation in which the energy dispersion of the hexagonal lattice is linearized around one of the two Dirac points where the valence and conduction band touch. But corrections to this approximation have to be included to discuss e.g. the recently measured absorption of suspended graphene in the visible-optics regime\cite{Nair08}, which is related to polarizability via the continuity equation. This has been done in a perturbative treatment in Ref.\cite{Stauber08}. The optical properties of graphite were calculated in Ref. \cite{Pedersen03}. 

Here, we want to extend the previous calculations to the full Brillouin zone of the hexagonal tight-binding model. We certainly expect deviations at large energies where the Dirac-approximation does not hold anymore. But the main purpose is to test whether the diverging density of states at the $M$-point (van Hove singularity) leads to consequences on the collective excitations of this system. 

Our interest is motivated by the recent findings of an additional plasmon mode that emerges at around $4.7$eV with a linear dispersion \cite{Valerio08} which was observed on freestanding graphene by electron energy loss spectroscopy \cite{Bangert08}. A first guess is to associate this mode to the van Hove singularity which in the charge susceptibility shows up at  $2t\approx5.4$eV, $t$ denoting the tight-binding hopping parameter. Including excitonic effects, the prominent absorption peak shifts to $4.5$eV \cite{Louie09}, thus suggesting that the van Hove singularity is indeed the origin for this new plasmon mode.

Apart from the high-energy corrections stemming from interband transitions, we also look at the intraband contribution at low energies and find that even there deviations from the Dirac cone approximation occur. By this, we complement a recent work, where the plasmon dispersion is discussed by also considering the full Brillouin zone \cite{Ziegler09}.

The paper is organized as follows. In Sec. II, we introduce the model and notation and define the polarizability of graphene. In Sec. III, we discuss the imaginary part of the polarizability which will only involve one numerical integration. We first treat the interband contribution where we especially focus on the behavior around the M-point where the van Hove singularity occurs. We then discuss the intraband contribution and the different behavior at certain directions of the incoming wave vector $\q$. In Sec. IV, we obtain the real part of the polarizability via the Kramers-Kronig relation and discuss implications on the modified plasmon spectrum due to the inclusion of the whole Brillouin zone. We close with conclusions in Sec. V and give details on the analytical evaluation of the polarizability around the M-point in an appendix.      

\section{The effective model and the polarizability of graphene}
\label{sec_model}
The Hamiltonian of a hexagonal graphene sheet in Bloch spinor representation is given by
\begin{align}
H=\sum_\k (-tH_\k-\mu1_{2\times2})\;,\; H_\k=\psi_\k^\dagger\begin{pmatrix}
0&\phi_\k\\
\phi_\k^*&0
\end{pmatrix}
\psi_\k\;,
\end{align} 
with $t\approx2.7$eV the tight-binding hopping parameter, $\mu$ the chemical potential and $\psi_\k=(a_\k,b_\k)^T$, $a_\k$ and $b_\k$ being the destruction operators of the Bloch states of the two triangular sublattices, respectively. Further, we have $\phi_\k=\sum_{{\bm \delta}_i}e^{-i({\bm \delta}_i-{\bm \delta}_3)\cdot\k}$ , where ${\bm \delta}_i$ denote the three nearest neighbor vectors. Here we choose them to be 
\begin{align}
{\bm \delta}_1=\frac{a}{2}(-1,\sqrt{3})\;,\;
{\bm \delta}_2=\frac{a}{2}(-1,-\sqrt{3})\;,\;
{\bm \delta}_3=a(1,0)\;,
\end{align}
with $a=1.42$\AA being the nearest carbon-carbon distance. The Brillouin zone is then defined by the two vectors ${\bf b}_{1}=2\pi/(3a)(1,\sqrt 3)$ and ${\bf b}_{2}=2\pi/(3a)(1,-\sqrt 3)$, see Fig. \ref{fig:bz} a). The dimensionless eigenenergies thus read
\begin{align}
|\phi_\k|=\sqrt{3+2\cos(\sqrt{3}k_ya)+4\cos(\sqrt{3}k_ya/2)\cos(3k_xa/2)}\;.
\end{align} 

In terms of the bosonic Matsubara frequencies $\omega_n=2\pi n/\beta$ ($\beta=1/k_BT$, $\hbar=1$), the polarizability in first order is defined as
\begin{align}
P^{(1)}(\q,i\omega_n)&=\frac{1}{A}\int_0^{\beta} d\tau e^{i\omega_n\tau}\langle \rho(\q,\tau)\rho(-\q,0)\rangle
\end{align}
where $A$ denotes the area of the graphene sample and the density operator is defined as $\rho_{\q}=\rho_\q^a+e^{i\q\cdot{\bm \delta}_3}\rho_{\q}^b$ with $\rho_\q^c=\sum_{\k,\sigma} c_{\k,\sigma}^\dag c_{\k+\q,\sigma}$ ($c=a,b$). 

Hence, we obtain the general expression for the polarizability
\begin{align}
\label{Polarization}
P^{(1)}(\q,i\omega_n)&=\frac{-g_s}{(2\pi)^2}\int_{\text{1.BZ}}d^2k\sum_{s,s'=\pm}f_{s\cdot s'}(\k,\q)\\\nonumber
&\times\frac{n_F(E^{s}({\bf k}))-n_F(E^{s'}({\bf k}+{\bf q}))}{E^{s}({\bf k})-E^{s'}({\bf k}+{\bf q})+i\omega_n}\;,
\end{align}
with $E^{\pm}({\bf k})=\pm t|\phi_\k|-\mu$ and $n_F(E)$ the Fermi function. For the band-overlap, we have
\begin{align}
f_{\pm}(\k,\q)&=\frac{1}{2}\left(1\pm\text{Re}\left[e^{i\q\cdot{\bm \delta}_3}\frac{\phi_\k}{|\phi_\k|}\frac{\phi_{\k+\q}^*}{|\phi_{\k+\q}|}\right]\right)\;.
\end{align}
Note that since we are summing over the entire Brillouin zone, only the spin-degeneracy $g_s=2$ has to be taken into account.

For neutral graphene, $\mu=0$, there is no intraband contribution due to the canceling Fermi functions in the numerator of Eq. (\ref{Polarization}). Due to $f_{-}(\k,\q\rightarrow0)\rightarrow0$, we further expect no interband contribution for $\q=0$. We finally note that for high energies $\omega>t$, the phase factor between the particle densities of the two sublattices, $e^{i\q\cdot{\bm \delta}_3}$, is crucial even in the long-wavelength limit $\q\rightarrow0$.

\section{Imaginary part of the polarizability}
With the substitution $i\omega_n\rightarrow\omega+i0$, the imaginary part of the retarded susceptibility is written in terms of a delta function in the usual way. Determining the zeros of the argument of the delta function allows to perform the integration over $k_x$ analytically. The subsequent integration over $k_y$ is then done numerically. We have also performed the direct summation of Eq. (\ref{Polarization}) of a finite system to check our results.
 
\begin{figure}[t]
\begin{center}
\includegraphics[angle=0,width=0.8\linewidth]{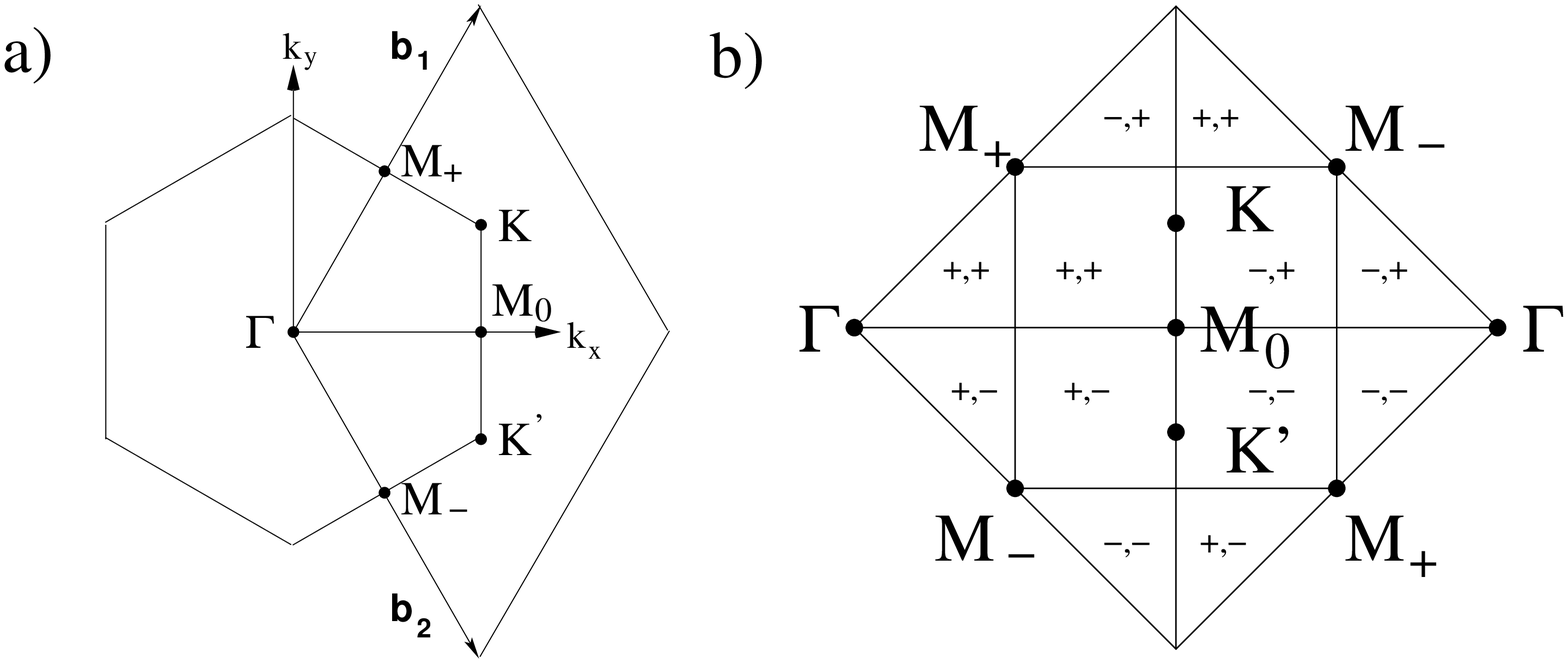}
\caption{a) The hexagonal and rhombical Brillouin zone b) The symmetrized rhombical Brillouin zone and its segmentation. The inner square refers to $j=-$, the outer triangles refer to $j=+$; additionally the values of $s$ and $s'$ are given.} 
  \label{fig:bz}
\end{center}
\end{figure} 
The integration over the Brillouin zone can be split up into separate parts with slight modifications of the integrand (see below and Fig. \ref{fig:bz} b)).
The final domain is then given by $0<3k_xa/2<\pi/2$ and  $0<\sqrt{3}k_ya/2<\pi/2$ and the substitution $x=\sin(3k_xa/2)$ and $y=\sin(\sqrt{3}k_ya/2)$ can be performed. The resulting expression (see Eq. (\ref{ImPol})) explicitly displays the inversion symmetry of $\q$ with respect to the $q_x$- and $q_y$-axis. The polarizability $P^{(1)}(\q,\omega)$ is also invariant under rotation of $\pi/3$, displaying the underlying lattice symmetry. We thus find the following symmetry:
\begin{align}
\label{HexSymmetry}
P^{(1)}(|\q|,\pi/6+\tilde\varphi,\omega)=P^{(1)}(|\q|,\pi/6-\tilde\varphi,\omega)
\end{align}
The subsequent plots thus only show four representative curves with $0\leq\varphi\leq\pi/6$.

\subsection{Interband transitions}
We shall first discuss the contribution of the interband transitions to the imaginary part of the polarizability. As explained above, we first perform the integral over $k_x\rightarrow x$, thus eliminating the delta function. For neutral graphene, $\mu=0$, at zero temperature $T=0$ this yields the following expression:

\begin{figure}[t]
\begin{center}
\includegraphics[angle=0,width=0.8\linewidth]{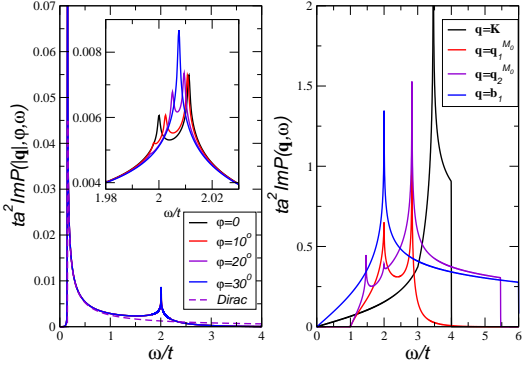}
\caption{(color online): Left hand side: The imaginary part of the polarizability $\text{Im} P^{(1)}(|\q|,\varphi,\omega)$ as function of the energy $\omega$ at $k_BT/t=0.01$ for different angles $\varphi$  with $|\q|a=0.1$. The result obtained from the Dirac-cone approximation is also shown (dashed line). Inset: Energy region around the van Hove singularity of the same curves. Right hand side: The imaginary part of the polarizability $\text{Im} P^{(1)}(\q,\omega)$ as function of the energy $\omega$ at $k_BT/t=0.01$ for various wave vectors $\q$ defined in the text.} 
  \label{fig:ImP}
\end{center}
\end{figure}

\begin{align}
\label{ImPol}
&\text{Im}P^{(1)}(\q,\omega)=\frac{2\text{sgn}(\omega)}{(2\pi)^2}\frac{\pi\sqrt{3}}{t}\left(\frac{2}{3a}\right)^2\int_0^{1} \frac{dy}{\sqrt{1-y^2}}\\\notag
&\times\sum_{j=\pm}\sum_{s,s'=\pm}\sum_{0<x_i<1}\frac{F_j(x_i,y;sq_x,s'q_y)}{\sqrt{1-x_i^2}|\frac{d}{dx}h_j(x,y;sq_x,s'q_y)|_{x_i}|}\quad,
\end{align} 
where we defined 
\begin{align}
\label{InterBand}
h_j(x,y;q_x,q_y)=|\phi_j(x,y;0,0)|+|\phi_j(x,y;q_x,q_y)|-|\omega|
\end{align}
with
\begin{align}
&|\phi_j(x,y;q_x,q_y)|=\Big[3+2(1-y^2)\cos(2\tilde{q}_y)\\\notag
&-4\sqrt{1-y^2}y\sin(2\tilde{q}_y)
+j4(\sqrt{1-x^2}\cos(\tilde{q}_x)\\\notag
&-x\sin(\tilde{q}_x))
(\sqrt{1-y^2}\cos(\tilde{q}_y)-y\sin(\tilde{q}_y))\Big]^{1/2}
\end{align}
and
\begin{align}
F_j(x,y;q_x,q_y)=\frac{1}{2}\left(1-\frac{\widetilde{F}_j(x,y;q_x,q_y)}{|\phi(x,y;0,0)||\phi(x,y;q_x,q_y)|}\right)
\end{align}
with
\begin{align}
&\widetilde{F}_j(x,y;q_x,q_y)=\cos(2\tilde{q}_x/3)+j2\sqrt{1-y^2}\\\notag
&\times\Big(\sqrt{1-x^2}\cos(2\tilde{q}_x/3)-x\sin(2\tilde{q}_x/3)\Big)\\\notag
&+2\Big(2\sqrt{1-y^2}\cos(\tilde{q}_x/3)+j\Big[\sqrt{1-x^2}\cos(\tilde{q}_x/3)\\\notag
&-x\sin(\tilde{q}_x/3)\Big]\Big)\left(\sqrt{1-y^2}\cos(\tilde{q}_y)-y\sin(\tilde{q}_y)\right)\;.
\end{align}
Above, we also introduced $\tilde{q}_x=3q_xa/2$ and  $\tilde{q}_y=\sqrt{3}q_ya/2$. Furthermore, the sum over $x_i$ is over all zeros which satisfy 
\begin{align}
h_j(x_i,y;q_x,q_y)=0
\end{align}
which can be written as a polynomial of fourth order. The zeros $x_i$ can thus be obtained analytically such that only the subsequent integration over $y$ has to be performed numerically.

On the left hand side of Fig. \ref{fig:ImP}, the imaginary part of the polarizability $\text{Im} P^{(1)}(|\q|,\varphi,\,\omega)$ as function of the energy $\omega$ is shown for different directions of the incoming wave vector $\q$ with $|\q|a=0.1$, where the usual parametrization in terms of the polar angle $\varphi$ with $q_x=|\q|\cos\varphi$ is used. There is no apparent angle dependence except for the region around the van Hove singularity which is highlighted in the inset. The result obtained from the Dirac cone approximation is also shown (dashed line), which is given by\cite{Gon94}
\begin{align}
\label{ImDirac}
\text{Im} P_{0,\rm Dirac}^{(1)}(|\q|,\omega)=\frac{1}{4}\frac{|\q|^2}{\sqrt{\omega^2-(3t|\q|a/2)^2}}\;.
\end{align}
For low energies, there is good agreement with the above formula, but especially for energies close to the van Hove singularity, $\omega=2t$, strong deviations are seen which shall be discussed in the following in more detail.

\subsubsection{Expansion around the van Hove singularity}
The new feature compared to the Dirac cone approximation comes from the region around the van Hove singularity, located at the $M$-points of the Brillouin zone. For the Brillouin zone defined above, the $M$-points are located at $M_0=2\pi/(3a)(1,0)$ and $M_\pm=\pi/(3a)(1,\pm\sqrt 3)$. 

In the following, we introduce the substitutions $\tilde{p}_x=3p_xa/2$ and  $\tilde{p}_y=\sqrt{3}p_ya/2$ and shall assume $\tilde{p}_x,\tilde{p}_y\ll1$ for $p=k,q$. Expanding around the $M_0$-point, the dispersion then simplifies to 
\begin{align}
\phi_{\k}^{M_0}&\approx-1-i2\tilde{k}_x+\tilde{k}_x^2+\tilde{k}_y^2\,,\\
\label{SaddlePoint}
|\phi_\k^{M_0}|&\approx1+\tilde{k}_x^2-\tilde{k}_y^2\;,
\end{align}
and the band-overlap yields
\begin{align}
\label{fM0}
f_{-}^{M_0}(\k,\q)\approx 4\tilde{q}_x^2/9\quad.
\end{align}
For the $M_\pm$-point, we obtain
\begin{align}
\phi_\k^{M_\pm}&\approx1\mp i2\tilde{k}_y\pm2\tilde{k}_x\tilde{k}_y\;,\\
|\phi_\k^{M_\pm}|&\approx1\pm2\tilde{k}_x\tilde{k}_y+2\tilde{k}_y^2\;,
\end{align}
and the band-overlap yields
\begin{align}
\label{fMpm}
f_{-}^{M_\pm}(\k,\q)\approx(\tilde{q}_x/3\pm\tilde{q}_y)^2\quad.
\end{align}

For $q_y=0$ ($\varphi=0$), an analytical approximation similar to that presented in Ref.\cite{Stauber07} is possible for the $M_0$-point expansion since the polynomial in the delta function is quadratic. This yields a logarithmic divergence at $\omega_M/t=\tw_M=2+\tilde{q}_x^2/2$ which can be approximated by  the following expression:
\begin{align}
\label{ImPM}
\text{Im}P^{(1),M_0}(q_x,\omega)&\approx\frac{2\text{sgn}(\omega)}{(2\pi)^2}\frac{\pi\sqrt{3}}{t}\left(\frac{2}{3a}\right)^2\frac{\tilde{q}_x^2}{18}\\\notag
&\times\left[\ln\left(\frac{8\Lambda^2}{\tilde{q}_x^2}\right)+\ln\left(\frac{2\Lambda^2\tilde{q}_x^2}{(\tw-\tw_M)^2}\right)\right]\;,
\end{align}
where $\Lambda$ denotes a suitable cutoff. Details on the calculation are given in the appendix. For general $\q$ or for the $M_\pm$-point expansion, we expect a similar behavior. 

In the inset of Fig. \ref{fig:ImP}, the region around the van Hove singularity which is highlighted. For a general angle $\varphi$, all three $M$-points contribute and there is a prominent double or even triple peak structure. But for $\varphi=\pi/6,\pi/2,..$ the overlap function of one $M$-point vanishes and the peaks merge.
 
\subsubsection{Behavior at large ${\bf q}$}

Let us now discuss the behavior for general ${\bf q}$ at $\mu=0$. For that we expand the energy dispersion $|\phi_{\k+\q}|$ around the points of high symmetry $\k=S=\Gamma,K,M$ and determine the ${\bf q}$-vector for which $|\phi_{\q}^{S}|=0$. To discuss the dielectric function at large wave-vectors $|\q|\sim1/a$, also local field effects have to be taken into account,\cite{Adler} as was done in Ref. \cite{Tudorovskiy09}.

Expanding the dispersion around $\Gamma=(0,0)$ and $K=2\pi/(3a)(1,1/\sqrt 3)$, we find that for ${\bf q}=K$, $|\phi_{\q}^{\Gamma,K}|=0$. The spectrum of $\text{Im}P^{(1)}$ thus starts at $\omega=0$. Expanding the dispersion around the $M_0$-point, the wave vectors $\q_1^{M_0}=2\pi/(3a)(0,1/\sqrt 3)$ and $\q_2^{M_0}=2\pi/(3a)(1,2/\sqrt 3)$ yield $|\phi_{\q}^{M_0}|=0$ and the spectrum of $\text{Im}P^{(1)}$ thus starts at $\omega=t$. For the $M_\pm$-point expansion, we obtain $|\phi_{\q}^{M_\pm}|=0$ for $\q=M_\mp$. 

On the right hand side of Fig. \ref{fig:ImP}, the imaginary part of the polarizability is shown for the above wave vectors $\q$. The curves for the $M_\pm$-point expansion yield the same curves as the ones for the $M_0$-point expansion and are not listed. For comparison, we also show the behavior for one of the vectors which define the Brillouin zone, $\q={\bf b}_{1}$, which (for $T=0$) is identical to the density of states by rescaling $\omega\rightarrow\omega/2$.

\subsection{Intraband transitions}

\begin{figure}[t]
\begin{center}
  \includegraphics[angle=0,width=0.8\linewidth]{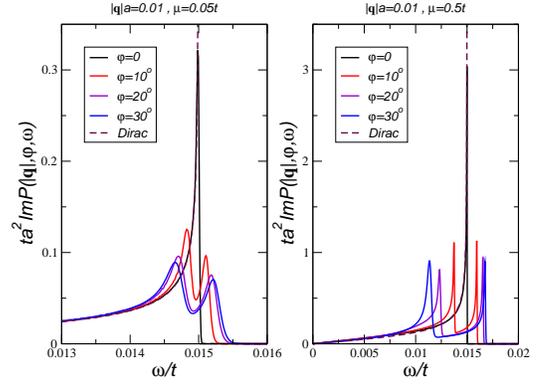}
\caption{(color online): The imaginary part of the polarizability $\text{Im} P^{(1)}(|\q|,\varphi,\omega)$ with $|\q|a=0.01$ as function of the energy $\omega$ for various angles $\varphi$ and two chemical potentials $\mu/t=0.05$ (left) and $\mu/t=0.5$ (right) at $k_BT/t=0.01$. Also shown is the analytical result coming from the Dirac cone approximation at zero temperature (dashed line).} 
  \label{fig:Intra1}
\end{center}
\end{figure}

For finite chemical potential $\mu>0$, there are also intraband
transitions.  The extension of Eq. (\ref{ImPol}) to finite chemical
potential $\mu$ and finite $T$ is straightforward. The main difference
is that the function of Eq. (\ref{InterBand}) now reads
\begin{align}
  \label{IntraBand}
  h_j(x,y;q_x,q_y)=|\phi_j(x,y;0,0)|-|\phi_j(x,y;q_x,q_y)|\pm|\omega|\;.
\end{align}
This expression suggests that there might be differences to the Dirac cone approximation also for small $\omega$. In fact, even for wave vectors $|\q|$ and chemical potential $\mu$ for which the Dirac cone approximation holds (e.g. $|\q|a=0.1$, $\mu/t=0.05$), we find deviations from the Dirac cone result.

Let us first start the discussion by summarizing the results coming from the Dirac cone approximation from Ref. \cite{Wunsch06} for which we introduce the functions 
\begin{align}
f(|\q|,\omega )& =\frac{1}{4\pi }\frac{|\q|^{2}}{\sqrt{|\omega ^{2}-(3t|\q|a/2)^{2}|}}\,,
\notag \\
G_{>}(x)& =x\sqrt{x^{2}-1}-\cosh ^{-1}(x)\,, \quad x>1 \, , \notag \\
G_{<}(x)& =x\sqrt{1-x^{2}}-\cos ^{-1}(x)\, , \quad |x|<1\, .
\label{eq:realG}
\end{align}
Due to the finite chemical potential, the polarizability is now given by $\text{Im}P_{\mu,\rm Dirac}^{(1)}=\text{Im}P_{0,\rm Dirac}^{(1)}+\text{Im}\Delta P_{\mu,\rm Dirac}^{(1)}$ and with $\omega_{\q}^D=3t|\q|a/2$ the additional term reads 
\begin{align}
&\text{Im}\Delta P_{\mu,\rm Dirac}^{(1)}(|\q|,\omega )=f(|\q|,\omega )\times\nonumber\\
&\left\{
\begin{array}{ll}
G_{>}(\frac{2\mu +\omega }{\omega_{\q}^D})-G_{>}(\frac{2\mu -\omega }{\omega_{\q}^D}) & ,\omega<\omega_{\q}^D \wedge \omega<2\mu-\omega_{\q}^D
\\[1.5ex]
-\pi & ,\omega>\omega_{\q}^D \wedge \omega<2\mu-\omega_{\q}^D \\
G_{>}(\frac{2\mu +\omega }{\omega_{\q}^D}) & ,\omega<\omega_{\q}^D \wedge \omega>\omega_{\q}^D-2\mu\\
G_{<}(\frac{\omega -2\mu }{\omega_{\q}^D}) & ,\omega>\omega_{\q}^D \wedge\omega<\omega_{\q}^D+2\mu\\
0 & ,\text{ otherwise} 
\end{array}
\right.  \;.\label{PmuDirac}
\end{align}

A distinct signature of non-interacting 2D
electrons in graphene is a divergent behavior of the polarizability or
charge susceptibility at the threshold
for the excitation of electron-hole pairs at $\omega_{\q}^D$, see Eq. (\ref{ImDirac}). This divergence is also present for gated or doped graphene with $\mu>0$ and has been usually attributed to the absence of curvature in the spectrum. But even in the regime where the Dirac cone approximation does not hold, i.e., curvature in form of trigonal warping has to be taken into account, we find a divergent behavior at $\omega_{\q}^D$ for $\q=q_x$. For arbitrary direction, we find a peak splitting even in the Dirac cone regime.

Let us discuss the polarizability using the full Brillouin zone in more detail. As stated above, $P^{(1)}(\q,\omega)$ is invariant under rotation of $\pi/3$ - independent of the chemical potential and we also find the symmetry of Eq. (\ref{HexSymmetry}). The numerical results moreover suggest that for moderate chemical potential $\mu<t$ the angle-dependent polarizability can be described by a single function where the angle $\tilde\varphi$ only enters as parameter.  

In Fig. \ref{fig:Intra1}, $\text{Im} P^{(1)}(q,\varphi,\omega)$ with $|\q|a=0.01$ as function of the energy $\omega$ is shown for various angles $\varphi$ and two chemical potentials $\mu/t=0.05$ (left) and $\mu/t=0.5$ (right) at $k_BT/t=0.01$. Only in the direction of $q_x$, i.e., $\varphi=0$, there is agreement with the analytical result of Eq. (\ref{PmuDirac}) coming from the Dirac cone approximation at zero temperature. Interestingly, this is also the case for a large chemical potential $\mu/t=0.5$, where trigonal warping effects should come into play. For arbitrary direction, a double-peak structure appears even for small chemical potential $\mu/t=0.05$ for which the Dirac cone approximation should hold. 

\begin{figure}[t]
\begin{center}
  \includegraphics[angle=0,width=0.8\linewidth]{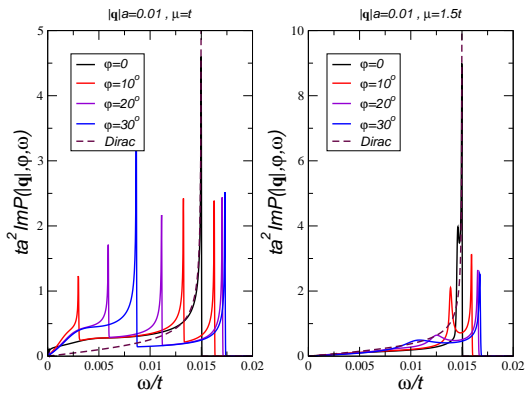}
\caption{(color online): The imaginary part of the polarizability $\text{Im} P^{(1)}(|\q|,\varphi,\omega)$ with $|\q|a=0.01$ as function of the energy $\omega$ for various angles $\varphi$ and two chemical potentials $\mu/t=1$ (left) and $\mu/t=1.5$ (right) at $k_BT/t=0.01$. Also shown is the analytical result coming from the Dirac cone approximation at zero temperature (dashed line).} 
  \label{fig:Intra2}
\end{center}
\end{figure}

In Fig. \ref{fig:Intra2}, the same quantities are shown for larger chemical potentials $\mu/t=1$ (left) and $\mu/t=1.5$ (right). The differences to the analytical result coming from the Dirac cone approximation (dashed line) now become apparent also for $\q=q_x$ ($\varphi=0$). For $\mu/t=1$ they manifest themselves at lower energies $\omega<\omega_\q^D$ and for $\mu/t=1.5$, a double-peak structure emerges. But interestingly, the divergence still occurs at $\omega\approx\omega_\q^D$ in both cases.  

The above curves were obtained for $k_BT/t=0.01$, thus slightly larger than room temperature, but we have also investigated the effect of different $T$. We find that the curves for $\varphi=0$ are basically unaffected by temperature, but that for arbitrary direction the algebraic divergences seen for $\mu/t=0.5$ or $\mu/t=1$ are smeared out at larger temperature as it is the case for $\mu/t=0.05$. On the contrary, the curves for $\mu/t=0.05$ and $\varphi\neq0$ develop the algebraic divergence for decreasing temperature. Generally, we can say that the algebraic divergences become broadened when the energy set by the temperature is much larger that maximal peak-splitting at $\varphi=\pi/6$. Nevertheless, the peak splitting in directions of lower symmetry prevails also at elevated temperatures.

\section{Real part of the polarizability}
The real part of the polarizability shall be obtained numerically via the Kramers-Kronig relation 
\begin{align}
\text{Re}P^{(1)}(\q,\omega)=\frac{1}{\pi}\int_0^{6t}d\omega'\text{Im}P^{(1)}(\q,\omega')\frac{2\omega'}{{\omega'}^2-\omega^2}\;.
\end{align}

\begin{figure}[t]
\begin{center}
  \includegraphics[angle=0,width=0.8\linewidth]{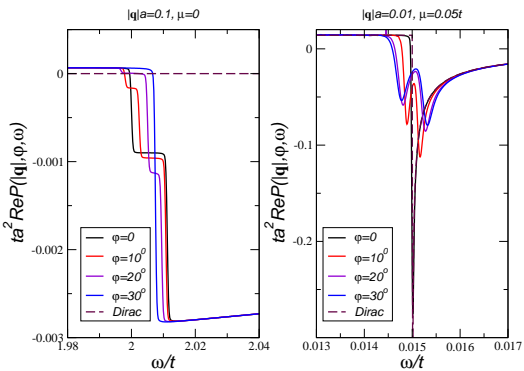}
\caption{(color online): The real part of the polarizability $\text{Re} P^{(1)}(|\q|,\varphi,\omega)$ as function of the energy $\omega$ for various angles $\varphi$ at $k_BT/t=0.01$. Left: For energies close to the van Hove singularity with $|\q|a=0.1$ and $\mu=0$. Right: For low energies with $|\q|a=0.01$ and $\mu/t=0.05$. Also shown is the analytical result coming from the Dirac cone approximation at zero temperature (dashed line).} 
  \label{fig:ReP}
\end{center}
\end{figure}

The left hand side of Fig. \ref{fig:ReP} shows $\text{Re} P^{(1)}(|\q|,\varphi,\omega)$ for energies close to the van Hove singularity with $|\q|a=0.1$ and $\mu=0$ for different angles $\varphi$. As expected, there are strong deviations with respect to the result coming from the Dirac cone approximation and the functions become negative. This opens up the possibility of the emergence of an additional plasmon mode since the plasmon dispersion in the RPA-approximation is given by the relation 
\begin{align}
\epsilon_\infty+v_{\q}P^{(1)}(\q,\omega)=0\;,
\end{align}
where $\epsilon_\infty$ denotes the effective dielectric constant including high-energy screening processes. Since the experiments in Ref. \cite{Bangert08} were done on suspended graphene, we set $\epsilon_\infty=1$. For the Coulomb interaction, we set $|\q|v_{\q}=e^2/2\varepsilon_0=90$eV\AA$\approx16\hbar v_F\approx24ta$. For $|\q|a=0.1$, $ta^2/v_{\q}=0.004$ never crosses one of the several curves which all tend to zero for larger energies. With the bare hopping amplitude $t\approx2.7$eV, we do thus not find an additional pole in the RPA-susceptibility.

Still, there is a renormalization of the hopping amplitude which comes from the wave function renormalization of the $\pi$ electrons. Near the van Hove singularity, this renormalization will be large and the matrix element will be reduced. With a renormalization of $t\rightarrow2t/3$, we would find an additional pole in the plasmon dispersion, consistent with experiments. 

\begin{figure}[t]
\begin{center}
  \includegraphics[angle=0,width=0.8\linewidth]{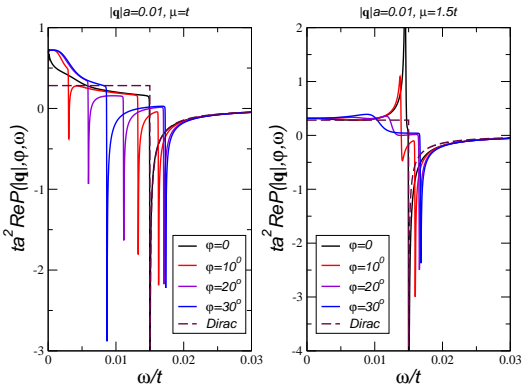}
\caption{(color online): The real part of the polarizability $\text{Re} P^{(1)}(|\q|,\varphi,\omega)$ for $|\q|a=0.01$ as function of the energy $\omega$ for various angles $\varphi$ at $k_BT/t=0.01$. Left: For the chemical potential $\mu/t=1$. Right: For the chemical potential $\mu/t=1.5$. Also shown is the analytical result coming from the Dirac cone approximation at zero temperature (dashed line).} 
  \label{fig:ReP_LargeMu}
\end{center}
\end{figure}

We also expect deviations from the simple one-particle spectrum around $\omega=2t$, because expanding the effective screened Coulomb potential within the RPA approximation, we have
\begin{align}
\text{Im}\frac{v_{\q}}{\epsilon(\q,\omega)}&=\text{Im}\frac{v_{\q}}{\epsilon_0+v_{\q}P^{(1)}(\q,\omega)}\\
&\approx -(v_{\q}/\epsilon_0)^2\text{Im}P^{(1)}(\q,\omega)\;.
\end{align}
As can be seen from Eq. (\ref{ImPM}), there is a logarithmic divergence at $\omega_M/t=2+(3q_xa)^2/8$ even for $q_x\rightarrow0$ since the prefactor from the band-overlap is canceled by $v_{\q}^2$.

The right hand side of Fig. \ref{fig:ReP} shows the real part of the polarizability $\text{Re} P^{(1)}(|\q|,\varphi,\omega)$ with $|\q|a=0.01$ and $\mu/t=0.05$ for different angles $\varphi$. For these parameters, the Dirac cone approximation is supposed to hold, but strong deviations are seen for $\varphi>0$ as in the case of the imaginary part. This opens up the possibility of a modified plasmon dispersion as discussed in Ref. \cite{Ziegler09}. But for the present parameters, we do not find an additional zero in the RPA-dielectric function, i.e., $ta^2\epsilon_\infty/v_{\q}=0.0004\epsilon_\infty$ crosses all curves at the same energy (choosing, e.g., the high-frequency dielectric constant of silicon $\epsilon_\infty=2$). Nevertheless, for larger wave numbers $|\q|a\approx\mu/t$, deviations are seen, i.e., the plasmon dispersion is more strongly damped and eventually vanishes since the square-root singularity is smeared out.    

In Fig. \ref{fig:ReP_LargeMu}, the real part of the polarizability is shown for two large chemical potentials $\mu/t=1$ (left) and $\mu/t=1.5$ (right). As for the imaginary part, large deviations compared to the results coming from the Dirac cone approximation (dashed line) are seen. First, the static value $P^{(1)}(|\q|,\varphi,\omega=0)$ is larger than $ta^2\text{Re}P_{\mu,\rm Dirac}^{(1)}(|\q|,\omega=0)=\frac{8}{9\pi}\mu/t$. Since the static value of the polarizability enters in the expression of the screened Coulomb potential, it is independent of the polar angle $\varphi$, consistent with group theory. Second, there are additional zeros of the real part of the RPA-dielectric function, i.e., $ta^2\epsilon_\infty/v_{\q}=0.0004\epsilon_\infty$ (set e.g. $\epsilon_\infty\approx2$) crosses the curves at various energies different from the one associated with the Dirac cone approximation. Nevertheless, they lie in the region where $\text{Im} P^{(1)}$ is finite and the new plasmon modes are thus damped. 

For the present parameters, the undamped solution occurs at slightly larger energy compared to the Dirac cone approximation but is independent of the direction $\varphi$. For larger wave numbers $|\q|a\approx\mu/t$, stronger deviations are seen, i.e., the plasmon dispersion depends on $\varphi$, is more strongly damped and eventually vanishes.    

\section{Conclusions}
We discussed the polarizability of graphene using the full band structure of the $\pi$-electrons. We especially focused on the features around the van Hove singularity since they might be responsible for the newly found plasmon dispersion. We find that there are no plasmon modes coming from the van Hove singularity within the RPA-approximation with the bare hopping amplitude $t$. But with a renormalization of $t\rightarrow2t/3$, there are additional plasmon modes, consistent with experiment.\cite{Bangert08} We also find a logarithmic divergence of the imaginary part of the effective Coulomb interaction. This will lead to prominent electron-hole interactions as was recently found in Ref. \cite{Louie09}.

We also looked at the intraband contribution to the polarizability. For $\q$ in the $\Gamma-M$-direction, we find basic agreement with the results of the Dirac cone approximation even for rather large chemical potential $\mu\approx t/2$, i.e., when corrections to the linear Dirac spectrum are large. For arbitrary direction of the incoming wave vector $\q$, we surprisingly found strong deviations from the results coming from the Dirac cone approximation where a double-peak structure emerges. The peak splitting occurs for all values of $\mu>0$ and is a direct consequence of the energy dispersion. For fixed $\q$ and $\mu$, it is largest for $\varphi=\pi/2$ and tends to zero for $|\q|,\mu\to0$. As a consequence, the plasmon dispersion is more strongly damped for $|\q|a\approx\mu/t$ and eventually vanishes at larger temperatures since the square-root singularities are smeared out.

\section{Acknowledgments}
We thank F. Guinea for useful discussions. This work has been supported by FCT under the grants PTDC/FIS/64404/2006, PTDC/FIS/101434/2008 and by Deutsche Forschungsgemeinschaft via GRK 1570.

\section{Appendix}
For $q_y=0$ and using the saddle-point approximation of Eq. (\ref{SaddlePoint}), an analytical calculation of the imaginary part of the polarizability around the $M_0$-point is possible. The integral is preformed in polar coordinates and with $x=\cos\varphi$, we have to solve the following quadratic equation with respect to $x$:
\begin{align}
|\phi_\k^{M_0}|+\phi_{\k+q_x}^{M_0}|-\tilde\omega=4k^2x^2\pm2k\tilde{q}_xx+\epsilon=0
\end{align}
with $\epsilon=2(1-k^2-\tilde\omega/2+\tilde{q}_x^2)$, $\tw=|\w/t|$ and $\tilde{q}_x=3q_xa/2$. The integral over $\varphi$ eliminates  the delta function and yields 
\begin{align}
\text{Im}P^{(1),M_0}(q_x,\omega)&=\frac{g_s\text{sgn}(\omega)}{(2\pi)^2}\frac{\pi\sqrt{3}}{t}\left(\frac{2}{3a}\right)^2\frac{8}{9}\tilde{q}_x^2\\
&\times\Big[{\cal I}_+(\tilde{q}_x,\tw)+{\cal I}_-(\tilde{q}_x,\tw)\Big]\;,
\end{align}
where we included the spin-degeneracy $g_s=2$ and defined the following integrals:
\begin{equation}
{\cal I}_\pm(\tilde{q}_x,\tw)=\int_{{\cal D}_\pm(\tilde{q}_x,\tw)}dkI_\pm(k;\tilde{q}_x,\tw)\;,
\end{equation}
with 
\begin{align}
I_\pm(k;\tilde{q}_x,\tw)&=\frac{k}{\sqrt{8(k^2-k_{\rm min}^2)}}\\\notag
&\times\frac{1}{\sqrt{8(k^2-\xi^2)\pm2\tilde{q}_x\sqrt{8(k^2-k_{\rm min}^2)}}}\;,
\end{align}
and the integration domains
\begin{equation}
{\cal D}_+(\tilde{q}_x,\tw)=\left\{\begin{array}{l l}
\left[k_{\rm min},\Lambda\right]&;\;\tw<2+5\tilde{q}_x^2/8\\
\left[k_-,\Lambda\right]\;&;\;\tw>2+5\tilde{q}_x^2/8
\end{array}\right.
\end{equation}
and
\begin{equation}
{\cal D}_-(\tilde{q}_x,\tw)=\left\{\begin{array}{l l}
\left[k_{\rm min},\Lambda\right]&;\;\tw<2+\tilde{q}_x^2/2\\
\left[k_{\rm min},k_-\right]\cup\left[k_+,\Lambda\right] \;&;\;2+\tilde{q}_x^2/2<\tw\\
&\;\;< 2+5\tilde{q}_x^2/8\\
\left[k_+,\Lambda\right]&;\;\tw>2+5\tilde{q}_x^2/8
\end{array}\right.\;.
\end{equation}
We further defined $\xi^2=-\tilde{q}_x^2/4-1+\tw/2$, $k_{\rm min}^2=q^2/8-\xi^2$, $k_\pm=|\xi\pm \tilde{q}_x/2|$, and $\Lambda$ denotes a suitable cutoff.

The indefinite integral has the following solution:
\begin{align}
&\int dkI_\pm(k;\tilde{q}_x,\tw)=\frac{1}{8}\ln\Bigg|\pm \tilde{q}_x\\\notag
&+\sqrt{8(k^2-k_{\rm min}^2)}+\sqrt{8(k^2-\xi^2)\pm2\tilde{q}_x\sqrt{8(k^2-k_{\rm min}^2)}}\Bigg|
\end{align}
There is a logarithmic singularity at $\tw_M=2+\tilde{q}_x^2/2$. Expanding the above result around $\tw_M$ yields the simple expression
\begin{align}
\text{Im}P^{(1),M_0}(q_x,\omega)&\approx\frac{g_s\text{sgn}(\omega)}{(2\pi)^2}\frac{\pi\sqrt{3}}{t}\left(\frac{2}{3a}\right)^2\frac{\tilde{q}_x^2}{18}\\\notag
&\times\left[\ln\left(\frac{8\Lambda^2}{\tilde{q}_x^2}\right)+\ln\left(\frac{2\Lambda^2\tilde{q}_x^2}{(\tw-\tw_M)^2}\right)\right]\;.
\end{align}


\end{document}